\newif\ifaastex\aastextrue

\ifaastex
\documentclass[useAMS,usenatbib,usegraphicx]{mn2e}
\else
\documentclass{emulateapj}
\renewcommand{\epsscale}[1]{}
\fi

\ifaastex
\else
\let\jjdagger=\dagger
\let\jjddagger=\ddagger
\renewcommand{\dagger}{\ensuremath{\jjdagger}}
\renewcommand{\ddagger}{\ensuremath{\jjddagger}}
\newcommand{\pasa}{PASA}



\fi

\makeatletter
\newcommand\ion[2]{#1$\;${\small\rmfamily\@Roman{#2}}\relax}%
\makeatother

\newcommand{\kms}{\ensuremath{\mathrm{km\,s}^{-1}}}
\newcommand{\msun}{\ensuremath{\mathrm{M}_{\odot}}}

\newcommand{\Halpha}{H$\alpha$}
\newcommand{\halpha}{H$\alpha$}

\newcommand{\Htwo}{\ensuremath{\mathrm{H}_2}}

\newcommand{\feii}{[\ion{Fe}{2}]}

\newcommand{\dmss}[4]{#1$\arcdeg$ #2$\arcmin$ #3$\farcs$#4}

\newcommand{\um}{\ensuremath{\mu\mathrm{m}}}

\newcommand{\uwishtwopaper}{F2011}

\newcommand{\feplus}{Fe$^{+}$}

\newcommand{\altaffilmark}[1]{$^{#1}$}
\newcommand{\altaffiltext}[2]{$^{#1}$#2}

\newcommand\arcdeg{\mbox{$^\circ$}}

\newcommand{\epsscale}[1]{}
\newcommand{\lesssim}{\la}

\newcommand{\plotone}[1]{\includegraphics[width=84mm,clip=true]{#1}}
\newcommand{\plottwo}[1]{\includegraphics[width=170mm,clip=true]{#1}}

\newcommand{\refjnl}[1]{{#1}}
\newcommand\aj{\refjnl{AJ}}% Astronomical Journal 
\newcommand\araa{\refjnl{ARA\&A}}% Annual Review of Astron and Astrophys 
\newcommand\apj{\refjnl{ApJ}}% Astrophysical Journal 
\newcommand\apjl{\refjnl{ApJ}}% Astrophysical Journal, Letters 
\newcommand\apjs{\refjnl{ApJS}}% Astrophysical Journal, Supplement 
\newcommand\aap{\refjnl{A\&A}}% Astronomy and Astrophysics 
\newcommand\aaps{\refjnl{A\&AS}}% Astronomy and Astrophysics, Supplement 
\newcommand\mnras{\refjnl{MNRAS}}% Monthly Notices of the RAS 
\newcommand\pasp{\refjnl{PASP}}% Publications of the ASP 
\newcommand{\pasa}{PASA}

\newcommand{\acknowledgements}{\vspace{1cm}}

\title[UKIRT Widefield Infrared Survey for \feplus]{UKIRT Widefield Infrared Survey for \feplus}

\author[J.~J. Lee et al.]{%
Jae-Joon Lee\altaffilmark{1,2},
Bon-Chul Koo\altaffilmark{3},
Yong-Hyun Lee\altaffilmark{3},
Ho-Gyu Lee\altaffilmark{1}, 
\newauthor%
Jong-Ho Shinn\altaffilmark{1},
Hyun-Jeong Kim\altaffilmark{3},
Yesol Kim\altaffilmark{3},
Tae-Soo Pyo\altaffilmark{4},
\newauthor%
Dae-Sik Moon\altaffilmark{5},
Sung-Chul Yoon\altaffilmark{3},
Moo-Young Chun\altaffilmark{1},
Dirk Froebrich\altaffilmark{7},
\newauthor%
Chris J. Davis\altaffilmark{8},
Watson P. Varricatt\altaffilmark{9},
Jaemann Kyeong\altaffilmark{1},
Narae Hwang\altaffilmark{1},
\newauthor%
Byeong-Gon Park\altaffilmark{1},
Myung Gyoon Lee\altaffilmark{3},
Hyung Mok Lee\altaffilmark{3},
Masateru Ishiguro\altaffilmark{3}\\
\\
\altaffiltext{1}{Korea Astronomy and 
Space Science Institute, Daejeon 305-348, Korea}\\
\altaffiltext{2}{leejjoon@kasi.re.kr}\\
\altaffiltext{3}{Department of Physics and Astronomy, Seoul National University, Seoul 151-747, Korea}\\
\altaffiltext{4}{Subaru Telescope, National Astronomical Observatory of Japan, 650 North A’ohoku Place, Hilo, HI 96720, USA}\\
\altaffiltext{5}{University of Toronto, Toronto, ON M5S 3H4, Canada}\\
\altaffiltext{6}{Visiting Brain Pool Scholar, Korea Astronomy 
     and Space Science Institute, Daejeon 305-348, Korea}\\
\altaffiltext{7}{Centre for Astrophysics \& Planetary Science, The University of Kent, Canterbury, CT2 7NH, UK}\\
\altaffiltext{8}{Astrophysics Research Institute, Liverpool John Moores University, UK}\\
\altaffiltext{9}{Joint Astronomy Centre, 660 North A’ohoku Place, University Park, Hilo, HI 96720, USA}
}

\begin{document}

\date{Accepted 2014 MMMM DD. Received 2014 MMMM DD; in original form 2014 MMMM DD}

\maketitle

\begin{abstract}

  The United Kingdom Infrared Telescope (UKIRT) Widefield Infrared
  Survey for Fe$^{+}$ (UWIFE) is a 180 deg$^2$ imaging survey of the
  first Galactic quadrant 
  ($7\arcdeg < l < 62\arcdeg$; $|b|   \lesssim   1\fdg5$) 
  using a narrow--band filter centered on the \feii\ 1.644 \um\ 
  emission line.
  The \feii\ 1.644 \um\ 
  emission is a good tracer of dense, shock--excited gas, and the
  survey will probe violent environments around stars: 
  star-forming regions, evolved stars, and supernova remnants, 
  among others. 
  The UWIFE survey is designed to complement the
  existing UKIRT Widefield Infrared Survey for \Htwo\
  \citep[UWISH2;][]{2011MNRAS.413..480F}. The survey will also
  complement existing broad-band surveys.
  The
  observed images have a nominal 5$\sigma$ detection limit of 18.7 mag
  for point sources, with the median seeing of 0.83\arcsec.
  For extended sources, we estimate surface brightness limit of
  $8.1 \times 10^{-20}$~W~m$^{-2}$~arcsec$^{-2}$ .
  In this   paper, we present the overview and preliminary results of this
  survey.

\end{abstract}


\section{Introduction}
\label{sec:intro}

\begin{figure*}
\epsscale{1.}
\plottwo{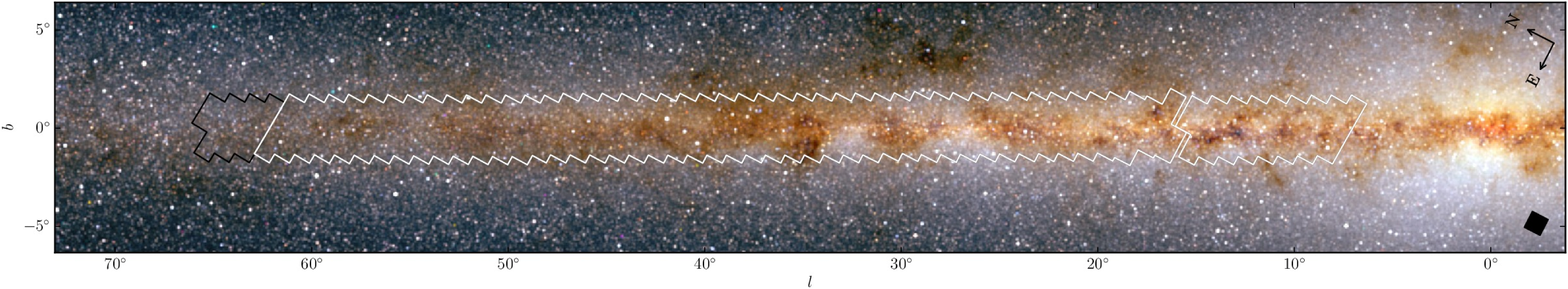}
\caption{An overview of the UWIFE survey area in the Galactic plane,
which are shown as the white polygons.
The background RGB composite image are the 2MASS K, H and J bands, 
respectively.
The black square in the lower right corner shows a nominal size and 
orientation of a single tile. 
The exact orientation of a tile will slightly change along the Galactic
plane as the orientation of the equatorial axis changes.
We note that the UWISH2 survey has completed its survey by 2011 and has
observed a total of 234 tiles (see texts for the definition of a tile), which
is slightly more than 224 tiles they
originally planned (\uwishtwopaper). The survey coverage we adopted is based
on their actual observations.
\label{fig:coverage-2mass}}
\end{figure*}

Stars, especially the massive ones, greatly influence the interstellar
medium (ISM) around them. The influence is not only due to the 
stellar radiation but also often via dynamical interaction, e.g.,
jets and outflows from protostars and mass-loss from evolved stars.
One of the most extreme case is their explosion as supernovae (SNe).
They heat and excite the interstellar gas, and also provide turbulent motions 
in the ISM. `Mass return' from stars to the ISM is particularly important
as it contributes to metal enrichment of the universe. 
Study of how the stars interact with their surrounding is thus 
of crucial importance not only to understand
the formation and evolution of stars but also to understand the evolution of
galaxies.

Emission lines are great probes of processes associated with the birth
and death of stars. The most popular
emission lines would be hydrogen Balmer lines in the optical.  The Balmer
lines are often emitted by gas ionized by young hot stars, thus are
used as a measure of the current star formation rate of galaxies
\citep{1998ARA&A..36..189K}.  The Balmer lines have played critical
roles for studying Galactic objects, and large scale imaging surveys
of the Galactic plane in \Halpha\ line 
\citep[e.g.,][]{2005MNRAS.362..689P,2005MNRAS.362..753D} have been a major 
resource for various studies.  The usage of Balmer lines, however, is  
limited for distant or embedded objects due to relatively high extinction
in the visual band.

With the advent of large--format infrared (IR) detectors, various efforts
of imaging surveys of lines in near--infrared (NIR) bands have been made.
The UKIRT Widefield
Infrared Survey for \Htwo\ \citep[UWISH2;][\uwishtwopaper\ hereafter]{2011MNRAS.413..480F}
is one such survey that probes
\Htwo\ 1--0 S(1) emission at 2.122 \um. 
Using the Wide-Field Camera onboard the 3.8-m UKIRT,
the survey covered the
northern portion of the Galactic plane of the %Spitzer Space Telescope
GLIMPSE survey \citep{2009PASP..121..213C}. 
\Htwo\ emission often traces outflows and jets from
embedded young stars and the regions around massive stars that are
radiatively excited.

Other prominent emission lines in NIR are lines from Fe$^+$. 
The Fe$^+$ ion has four ground terms, each of which has 3 -- 5
closely-spaced levels to form a 16 level system
\citep{2011aas..book.....P}. 
The energy gap between the ground level and the excited levels
is less than $1.3 \times 10^4$ K, making these levels easily excited
in the postshock cooling region. The emission lines resulting from
the transitions among them appear in optical- to far-infrared
bands. In the NIR JHK bands, 10--20 [Fe II] lines are visible where the
two strongest lines are at 1.257 and 1.644 \um\ 
\citep[]{2013arXiv1304.3882K}.

The
\feii\ lines are relatively strong in shocked gas rather than
photoionized gas because Fe atoms in photoionzed gas are likely in higher
ionization stages, and also because the Fe abundance in shocked
gas is probably enhanced by dust destruction \citep{2013arXiv1304.3882K}.  
Behind radiative atomic shocks, [Fe II] emissions
are mostly emitted from the cooling region.  For
example, %unless there is heavy-element-enriched SN ejecta
the ratio of \feii\ 1.644 \um\ to hydrogen Pa$\beta$ line in supernova
remnants (SNRs) is between 2 and 7, while it is as low as $0.013$ in 
starforming regions in Orion
\citep{1989A&A...214..307O,2000ApJ...528..186M}.
Therefore, [Fe II] forbidden lines could be used as
tracers of fast radiative atomic shocks, although they can be
emitted from photoionized gas when ionized by
high energy photons such as  X-rays,
e.g., in active galactic nuclei \citep{2000ApJ...528..186M}.

Here we present an unbiased survey of a portion of the Galactic Plane
in \feii\ 1.644 \um\ emission, dubbed as UKIRT Widefield Infrared
Survey for Fe$^{+}$ or UWIFE.  
The survey was aimed to detect ionized Fe objects
(IFOs), where the \feii\ line will highlight regions
of predominantly shock-excited dense gas. \feii\ emission is an
excellent tracer of dense outflows in young and massive star forming
regions, eruptive mass loss from evolved stars, and dense media
interacting with SNRs. 
The survey 
was designed to cover the same region and 
to complement the UWISH2 survey.
The UWIFE survey was
conducted with the same telescope and detector combination as
in the UWISH2 survey, but only with a different filter. Further, the
survey complement the UKIDSS Galactic Plane JHK survey
\citep[GPS;][]{2008MNRAS.391..136L}, and various other surveys like
the IPHAS \Halpha\ survey \citep{2005MNRAS.362..753D}, the VLA 5 Ghz
CORNISH survey \citep{Hoare(2012)PASP_124_939}, etc.

The UWIFE survey started its initial observations in the summer of 2012 and 
has been completed in September of 2013. In this paper, we define
the survey (\S\,\ref{sec:obs}) and describe
its characteristics (\S\,\ref{sec:results}). Our survey 
is briefly compared to \halpha\ surveys in \S\,\ref{sec:feii-char}.
We further outline the scientific objectives of the survey with some
preliminary results when available
(\S\,\ref{sec:discussion}). \S\,\ref{sec:conclusion} concludes the
paper.

\begin{figure*}
\epsscale{1.}
\plottwo{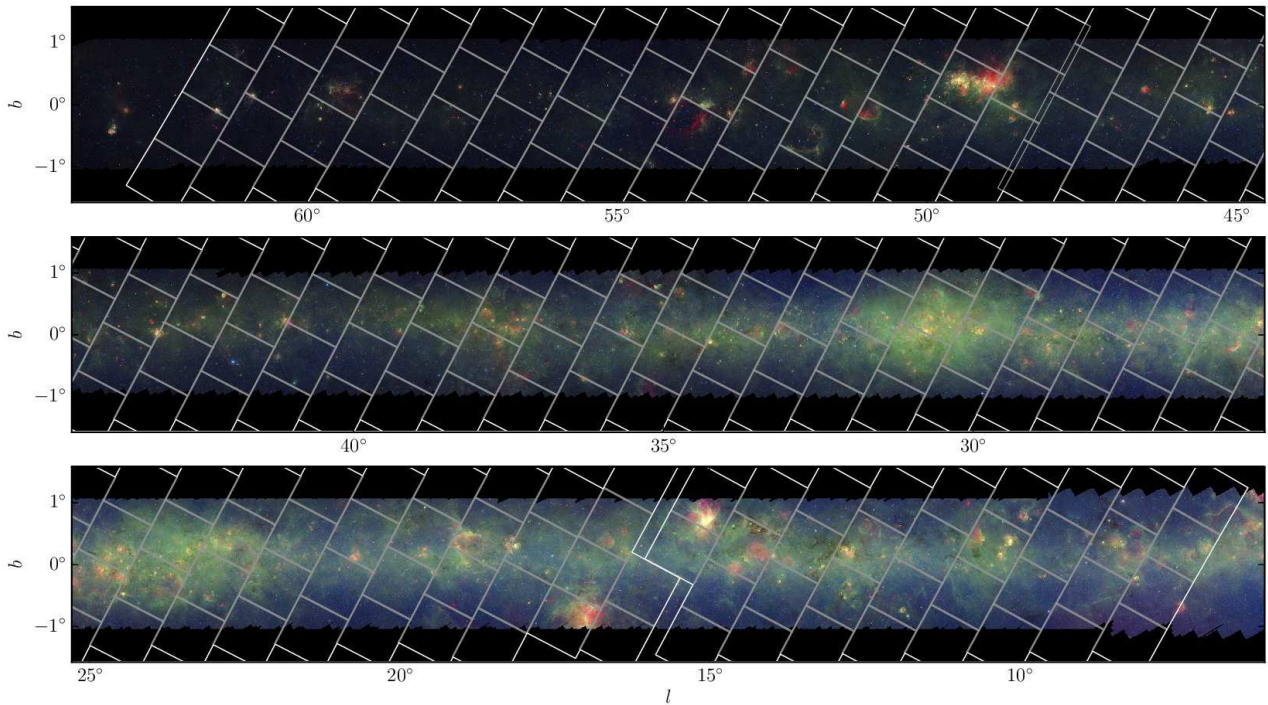}
\caption{Footprints of the UWIFE survey. The white boundary outlines the 
UWIFE survey area. The grey rectangles are coverage of individual tiles.
The background RGB composite image are from the Spitzer 
GLIMPSE and MIPSGAL survey.
Blue and green represents IRAC 3.6\,\um\  and 8\,\um, while red is MIPS\,24\um.
Note that there are slight gaps between adjacent tiles near 
$l\sim16\arcdeg$.
\label{fig:coverage-mipsgal}}
\end{figure*}

\section{TARGET AREA and OBSERVATION}
\label{sec:obs}

The UWIFE survey was conducted using the Wide-Field Camera
\citep[WFCAM,][]{2007A&A...467..777C} at 
UKIRT.
The camera has four Rockwell Hawaii-II HgCdTe
arrays. Each array has 2048 $\times$ 2048 pixels, corresponding to
$13.65\arcmin \times
13.65\arcmin$ field--of--view with a pixel scale of 0.4 arcsec. 
They are arranged in a square pattern, with a gaps 
of $12.83\arcmin$ between adjacent arrays.
With this
layout, observing at four discrete positions results in a contiguous
area covering 0.75 deg$^2$ on the sky (a WFCAM tile).  For each pointing,
images were obtained taken at three jitter positions. The jitter offsets were
($0\arcsec$, $0\arcsec$), ($6\farcs4$, $0\arcsec$) and ($6\farcs4$,
$6\farcs4$). 
About each jitter position, we performed a 2 $\times$ 2 microstep with
offset size of 
$4.62\arcsec$, to fully
sample the point spread function. Three jitter positions with $2
\times 2$ microstepping give total of 12 images.  An exposure time of
60 seconds was used, giving a total per-pixel integration time of 
720 seconds.
The final stacked images are resampled to 0.2\arcsec.  

The survey covered a region within the First Galactic
Quadrant ($7\arcdeg \lesssim l \lesssim 62\arcdeg$; $|b| \lesssim 1.5\arcdeg$ ).
The coverage is mostly identical to that of the UWISH2 survey except
that the UWISH2 covered up to $l \simeq 65\arcdeg$.
Figure~\ref{fig:coverage-2mass}
shows an overview of the UWIFE survey area in the Galactic plane
and compares it with the UWISH2 survey area. 
The
survey area consists of 220 tiles, where a single tile is a
square of $54\arcmin\times54\arcmin$, aligned in equatorial
coordinates. The tiles are arranged as 55 stripes of four consequent
tiles along lines of constant declination (see
Figure~\ref{fig:coverage-mipsgal}). The position of tiles are
identical to those of UWISH2, except 20 tiles in $43\arcdeg < l <
48\arcdeg$. For these
tiles, the coordinates of UWISH2 differ from those of the GPS by a few
arcminutes (they are identical otherwise), 
and we adopted the coordinates of the GPS.  
There are small gaps around $l\sim16\arcdeg$ not
covered by the survey. The gaps inherit from the survey area of UWISH2.
However, we
consider the impact of the gaps on the survey as minimal.

\section{Results}
\label{sec:results}

\subsection{\feii\ Filter}
\label{sec:feii-filter}

\begin{figure}
\epsscale{1.}
\plotone{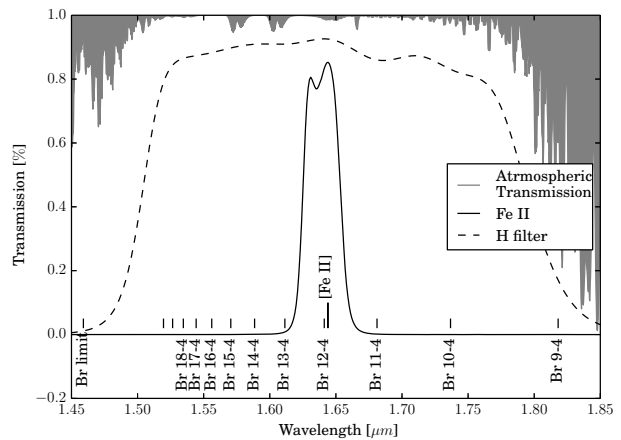}
\caption{The response curve of the narrow-band \feii\ 1.644 \um\ filter (solid
black), broad-band H (dashed black), and atmospheric transmission curves (gray
line). 
\label{fig:filter-response}}
\end{figure}

\begin{figure*}
\epsscale{1.}
\plottwo{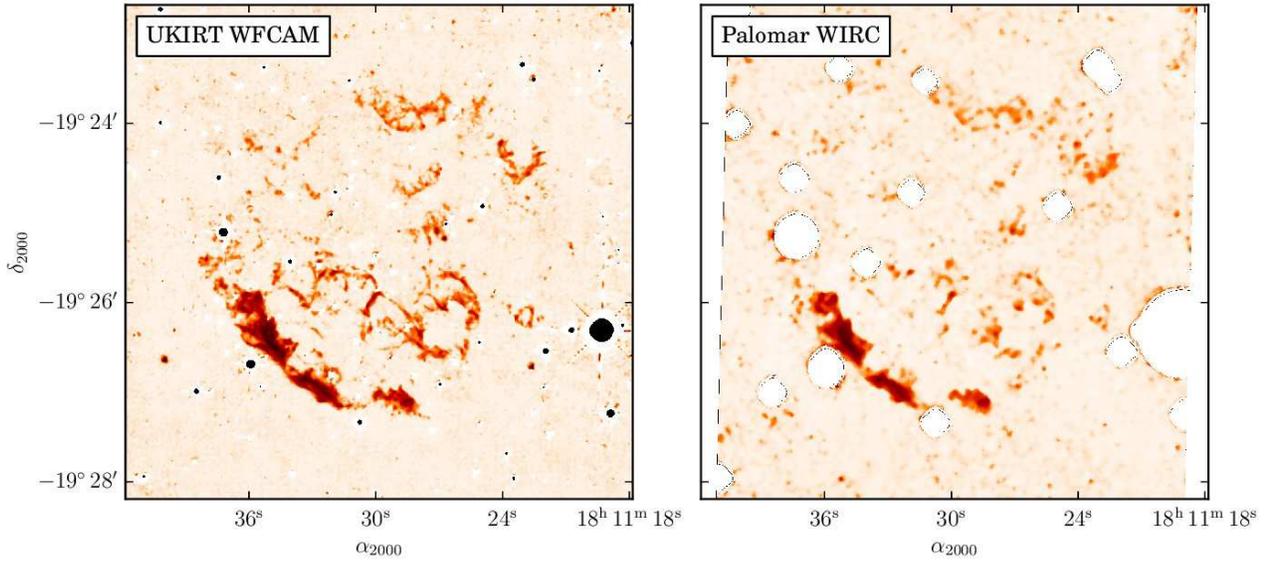}
\caption{\emph{(left)} UWIFE image of the SNR  G11.2$-$0.3, which
was taken with the new \feii\ 1.644 \um\ filter installed in WFCAM. 
\emph{(right)} Image of same region taken with 
\feii\ 1.644 \um\ filter on WIRC 
in Palomar Hale telescope
\citep{2007ApJ...657..308K}.
\label{fig:g11}}
\end{figure*}

The \feii\ 1.644~\um\ filters were not
available on WFCAM, so a set of new 
filters were procured from JDS Uniphase 
Corporation 
and installed during the
summer of 2012.  The filters have a central wavelength of 1.644~\um\ with
transmittance of 85\% and effective bandwidth of 0.026~\um.
Figure~\ref{fig:filter-response} plots the filter response curve
of the \feii\ filter provided by the manufacturer together with 
that of H band filter.
Figure~\ref{fig:g11} shows the first light image 
obtained using the newly installed filter, on the
SNR G11.2$-$0.3.  The image is
continuum-subtracted as described in
section~\,\ref{sec:results-subtraction}.  
The figure shows a comparison of our image
with the \feii\ image of the
same target obtained with the WIRC
onboard the Palomar Hale telescope \citep{2007ApJ...657..308K}. 
Our UKIRT WFCAM image
is basically identical to that of the WIRC, but shows more % finer
details that were not available in the WIRC image.

\subsection{Survey Status \& Data Quality}
\label{sec:results-comleteness}

\begin{figure*}
\epsscale{1.}
\plottwo{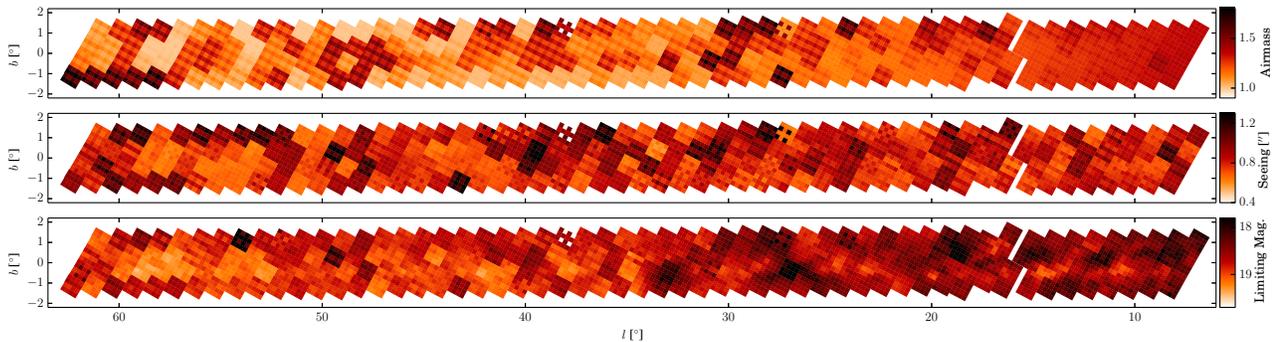}
\caption{Distributions of airmass, seeing, and limiting magnitude 
 of the obtained images at their sky positions. 
\label{fig:stat-skymap}}
\end{figure*}

\begin{figure}
\epsscale{1.}
\plotone{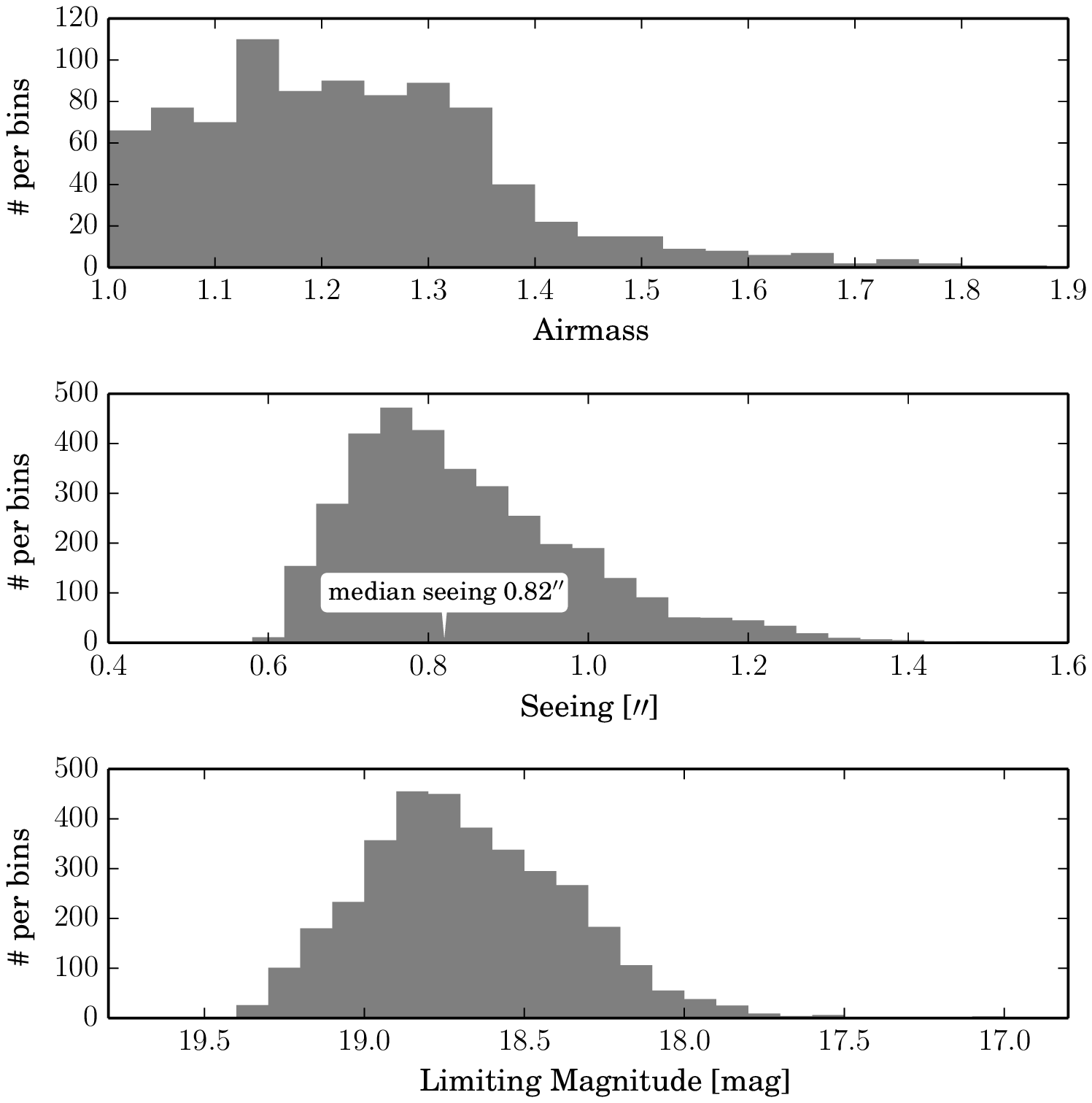}
\caption{Histogram of the airmass, seeing and limiting magnitude.
\label{fig:stat-hist}}
\end{figure}

\label{sec:results-quality}

The UWIFE observations were conducted during 
2012 and 2013 and we have observed total of
220 tiles.
The images were reduced by the Cambridge Astronomical 
Survey Unit (CASU), using the same pipeline used for 
reducing the images of UWISH2 and GPS.
The CASU reduction steps
are described in detail by 
\citet{2006MNRAS.372.1227D}; astrometric and
photometric calibrations \citep{2009MNRAS.394..675H} are done using
2MASS catalogue.

Figure~\ref{fig:stat-skymap} displays airmass, seeing, and limiting
magnitudes for 220 tiles  (one tile consists of 16 images).
The values are shown for
each image of a single chip. For
airmass values, 4 images in the same sequence share a same value,
resulting in a particular pattern in the plot. The seeing values are
obtained from the CASU pipeline products, and measured from 
the co-added frames. The statistics of these values are also
shown in Figure~\ref{fig:stat-hist} as histograms.
The majority of the data have a seeing between 0.6\arcsec and
1.0\arcsec. The median seeing is 0\farcs82, which is slightly worse than
that of UWISH2 survey (0\farcs73). Virtually all the images have seeing
better than 1\farcs5.

The 5$\sigma$ limiting
magnitudes are estimated following the definition of 
\citet{2006MNRAS.372.1227D}. %Dye et al. (2006).
\[
m = m0 - 0.05 * (X -1) - 2.5 \log_{10} \left(\frac{5 \sigma_{sky} (1.2N)^{1/2}}{t_{\mathrm{exp}}}\right) - m_{\mathrm{ap}}
\]
$m0$, $\sigma_{sky}$, and $m_{ap}$ are the zero point magnitude, sky
noise and the aperture correction value, which are obtained from the FITS header
of the pipeline products (Header keyword of MAGZPT, SKYNOISE and ACOR3).
$X$ is the airmass and N is the number of pixels in a given aperture
(aperture size of 1\arcsec\ is used corresponding to the header value
of APCOR3). The photometric calibration of the CASU pipeline is done with
the 2MASS point source catalog and H band magnitudes are used for
calibration of \feii\ images where the zero magnitude flux is
$1.133 \times 10^{-13}$ W cm$^{-2}$ $\um$$^{-1}$ \citep{2003AJ....126.1090C}.
The typical values of $m0$, $\sigma_{sky}$, and $m_{ap}$
are 22.25 mag, 20 counts, and 0.2 mag, respectively.  For most images,
the calculated limiting magnitudes are fainter than 18 mag, with a
modal value around 18.7.  However, we expect the real limiting
magnitudes to be brighter as crowding along the Galactic plane 
is not fully accounted when sky noise values are estimated. 
Similarly, from the $\sigma_{sky}$ values, we estimate a typical rms noise 
level of $8.1 \times
10^{-20}$~W~m$^{-2}$~arcsec$^{-2}$ for diffuse \feii\ emission, where we
used an effective filter bandwidth of 0.026 \um. Again, this would have been
underestimated in the inner Galactic regions.

\subsection{Continuum Subtraction}
\label{sec:results-subtraction}

\begin{figure}
\epsscale{1.}
\plotone{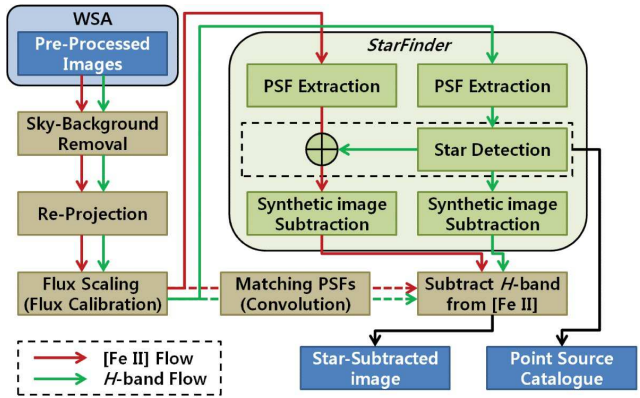}
\caption{The Flow chart of processes adopted for the conitnuum subtraction.
\label{fig:flowchart}}
\end{figure}

The observed \feii\ images are continuum-subtracted using the {\it H}
--band images from the GPS. The {\it H}--band images are
regridded and rescaled to match the astrometry and the flux scale of
the corresponding \feii\ images. The point--spread--function (PSF) of
\feii\ images and corresponding H images often differ significantly.
To compensate the effect of the different PSF, an image of a better
seeing can be convoluted to match the poorer seeing of the other image.
However, this approach is not optimal as the resulting difference image 
have PSF of the poorer seeing. This 
is an important issue for \feii\ emission since \feii\ emission is often 
knotty and compact. 
To keep the original seeing of the 
\feii\ emission, we adopted an alternative continuum
subtraction method where we subtract point-like sources from the images by
modeling the PSF. Below, we describe more
details of this method.
\begin{enumerate}
\item We first re-project the {\it H}-band image to the corresponding 
  \feii\ image so that the two images are
  astrometrically aligned. 
\item We extract a $51\times51$ pixel area around well-isolated stars
  whose fluxes are greater than $30\sigma$ above the background.
  They are normalized and median-averaged to form the PSF image.
  To compensate the spatial variation of PSF across the focal plane,
  A single chip area (4096 $\times$ 4096 pixels) is divided into
  sub-regions and PSFs are estimated in each sub-region, where
  we find $4\times4$ sub-regions are mostly adequate.
\item We perform PSF photometry. Among the detected sources, we reject
  sources that are not point-like (sources with a PSF correlation
  coefficient less than 0.7). The source catalogues of \feii\ and 
  {\it H}--band are matched and we further reject sources that have a
  brightness difference larger than 20\% and/or a position difference
  larger than 0.5 pixels. With the second step we distinguish between 
  continuum point sources and unresolved line emission features.
\item We reconstruct synthetic images using the source catalogue and
  the PSFs. These synthetic images are subtracted from the original
  images. This step is done for both \feii\ and {\it H}--band images, and it
  removes most of the point-like continuum sources from both images. 
\item We then subtract the point-sources-removed {\it H}--band images from the 
  point-sources-removed \feii\ images. This is to remove other
  extended continuum sources that escaped our detection in previous steps.
\end{enumerate}
Figure~\ref{fig:flowchart} shows the overall flowchart of 
our continuum--subtraction method.
We primarily used {\it Starfinder} \citep{2000A&AS..147..335D} for the
PSF photometry together with Scamp 
\citep{2006ASPC..351..112B}
and Swarp 
\citep{2002ASPC..281..228B}
for the astrometry and
the image reprojection.

We note that the difference image between \feii\ and {\it H}--band images
may have contamination other than \feii\ emission.
The GPS images are taken during 2005--2008, and
the difference may due to source variability 
over more than a few years of period, in particular for point sources.
It is also possible that there is some, but not significant, contributions 
from lines other than \feii\ 1.644\,\um, or a differing continuum slope. 
For example, the hydrogen Br~12
line is within the bandwidth of our \feii\ filter. However, we believe
its contribution in the continuum-subtracted images are not
significant as the H band images contain Br 10 -- 18 lines and a good
fraction of the Br~12 line is subtracted when H band images are subtracted.

\subsection{Data Release}
\label{sec:data-release}

All the images from the survey and the continuum-subtracted images
will be made public eventually through our survey web page
(http://gems0.kasi.re.kr/uwife). At the time of writing, all the CASU
pipeline products are available together with most of the
continuum-subtracted images. The web page also provides pre-generated
RGB composite images (UWISH2, UWIFE and GPS J for red, green and blue,
respectively) of the entire survey region that are zoomable down to
resolution of 0\farcs4.
% Using any of our data in a publication will be allowed without
% restriction as far as a proper citation to this paper is made
% in the publication., although we encourage prospect users to
% contact the survey team for collaboration.

\section{Comparison with \halpha\ Surveys}
\label{sec:feii-char}

\begin{figure}
\epsscale{1.}
\plotone{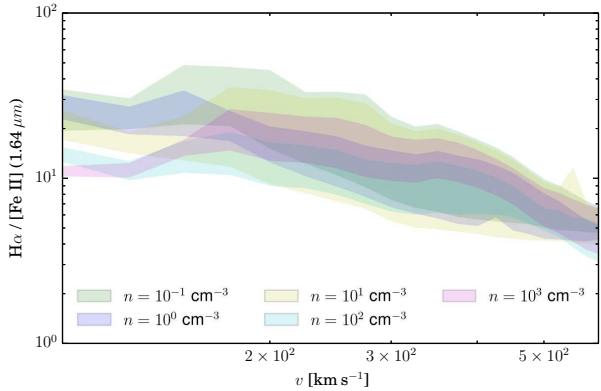}
\caption{
The ranges of model--calculated intrinsic line ratios of \feii\ 1.644\,\um\ 
to \halpha\ as a function of shock velocities from the shock grids of the Solar abundances from \citet{2008ApJS..178...20A}. 
Different colors 
represent different preshock densities. The vertical range of each plot
corresponds the range of magnetic parameters ($B/n^{1/2}$) from 
$10^{-4}$ to $10$ $\mu$G cm$^{3/2}$.
\label{fig:line-ratio}}
\end{figure}

\begin{figure}
\epsscale{1.}
\plotone{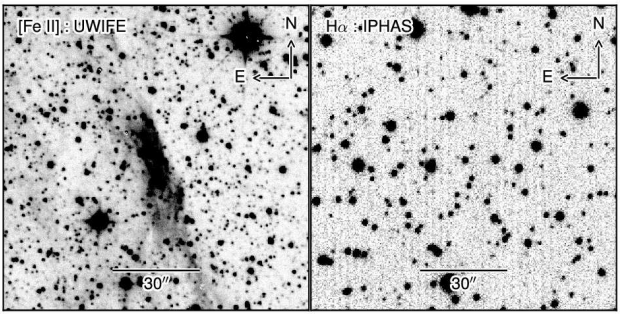}
\caption{
Northwestern part of SNR 3C391 centered at ($l$, $b$) = (\dmss{31}{52}{08}{7}, \dmss{+2}{07}{28}{5}). It shows bright \feii\ emission in UWIFE survey 
while \halpha\ emission from IPHAS survey is not detected.
Total hydrogen column densities toward 3C391 is estimated as $2 \times 10^{22}$ cm$^{−2}$, corresponding to a
visual extinction A$_{\mathrm V}$ $\sim$ 10 mag \citep{1996ApJ...467..698R}.
\label{fig:threecthreenineone-feii-ha}}
\end{figure}

The \feii\ forbidden lines, in particular the line at 1.644~\um, are  good
tracers of fast radiative atomic shocks. The intrinsic emissivity of
\feii\ line at 1.644 \um\ from atomic shocks could be fainter than
other tracers in optical bands (e.g., \halpha) in general. However, 
this can easily be compensated by lower
extinction in the infrared for embedded objects,
or those at large distances from us.

Figure~\ref{fig:line-ratio} shows model--calculated intrinsic line
ratios of \halpha\ to \feii\ 1.644\,\um\ for various shock conditions
assuming solar abundances adopted from the shock grids of
\citet{2008ApJS..178...20A}.  The models suggest that 
the \halpha\ emission is
intrinsically brighter than the \feii\ emission by an order of
magnitude for interstellar shocks of given shock velocity range 
of 100 to 500 \kms.
The ratio can become as high as $\sim50$ for slower shocks and as low as 
$\sim5$ for faster shocks.

The line ratios given above do not 
take the interstellar extinction into account.  Due to 
relatively lower extinction in the infrared, when
compared to that in the optical, the \feii\ lines
can appear brighter than \halpha\ for objects with high
foreground absorption.
We may define the amount of differential extinction
between \feii\ and \halpha\ for a given Av as
$10^{-2.5 \times (\mathrm{A}(1.644\mu m) - \mathrm{A}(0.656\mu m))}$.  
For example,
we get $\sim30$ for Av of 6 mag. 
This means that, 
although 
\feii\ is intrinsically fainter by a factor of $30$, \feii\ look as
bright as \halpha\ for Av of 6 mag.
According to Figure~\ref{fig:line-ratio}, the intrinsic line 
ratios of \halpha\ to \feii\ 1.644\,\um\ 
range from $5$ to $50$, and
it is expected that \feii\ becomes apparently
brighter than \halpha\ for objects with Av higher than several magnitudes.
The visual extinction of 6 mag is expected
for nominal hydrogen column density of $N_{\mathrm{H}} = 10^{22}$
cm$^{-2}$, thus \feii\ 1.644\,\um\ line would be often brighter than
\Halpha, particularly for objects in the inner galaxy. %  tic region.  

The above comparison does not take account the sensitivity of
the underlying detector system. Because of the high sky background in 
the NIR when compared to that in the optical, 
a better sensitivity is usually obtained at optical wavelengths,
although it depends on various other factors such as the telescope
diameter.  One of the surveys that well complements our \feii\ survey
is the INT/WFC Photometric \halpha\ Survey of the Northern Galactic
Plane (IPHAS), which is an \halpha\ imaging survey being carried out
using the 2.5m Isaac Newton Telescope. The IPHAS survey has an
\halpha\ surface brightness limit of
$10^{-20}$~W~m$^{-2}$~arcsec$^2$ \citep{2013MNRAS.431..279S}, which is
comparable to the surface brightness limit of our survey. Therefore,
the UWIFE survey will have advantage over IPHAS survey for sources
with Av greater than several magnitudes.
Figure~\ref{fig:threecthreenineone-feii-ha} shows one example
at A$_{\mathrm V}$ $\sim$ 10 mag where we see bright \feii\ emission
while \halpha\ emission is absent.

\section{Discussion}
\label{sec:discussion}

\subsection{Jets/Outflows from Young Stars}
\label{sec:discussion-sfr}

\label{sec:discussion-yso}

\begin{figure}
\epsscale{1.}
\plotone{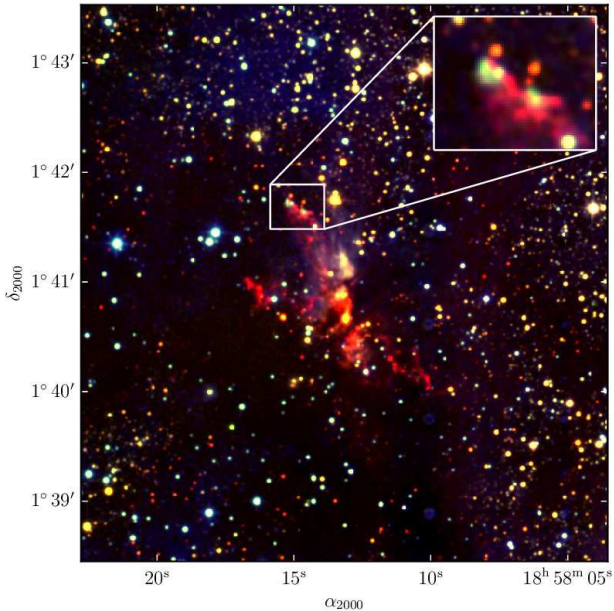}
\caption{RGB colour composite image of the star-forming region G35.2N.
Red, green and blue each represents narrow band \Htwo\ from the UWISH2, 
narrow band \feii\ from the UWIFE and broad--band J from the GPS.
Unlike broadly distributed \Htwo\ emission, \feii\ is shown
as compact emission at the tip of jets.
\label{fig:gthirtyfive}}
\end{figure}

Outflows are ubiquitous in low- to high-mass star formation
\citep{1996ApJ...472..225S}, so they are significant
signposts in searching for star forming regions.
Since the discovery of Herbig-Haro (HH) objects
\citep{1950ApJ...111...11H,1952ApJ...115..572H} in the optical, many
different types of outflows have been reported over a wide range of
wavelengths, from X-ray to radio
\citep{1985ARAA..23..267L,1996ARAA..34..111B,2001ARAA..39..403R,2005IAUS..227..237S,2007prpl.conf..215B}.
Each emission line in the various wavelength realms from X-ray to the
radio shows different aspects in the physical
and chemical conditions in the shocks.
In the near-infrared, \feii\ and H$_2$ emission lines are good
tracers for outflows and jets and
have often been observed in  low-mass star forming regions
\citep{1995AA...300..851D,2000AJ....120.1449R,2003AA...397..693D,2009ApJ...694..582H}.
Generally \feii\ emission traces partially ionized, fast, and
dissociative J-shocks, while \Htwo\ emission traces neutral, slow,
C-shocks or non-dissociative shocks
\citep{1989ApJ...342..306H,1994AA...289..256S}.
These two emission lines show different spatial distribution and
properties each other
as demonstrated in Figure\,\ref{fig:gthirtyfive}.
While the \feii\ emission is confined in narrow jets or relatively
compact knots at the tip of jets,
\Htwo\ emission can be more extended.
The fast well collimated \feii\ jets close to the driving sources
provide the clue of the launching machanism
\citep{2003ApJ...590..340P,2006ApJ...649..836P,2009ApJ...694..654P}.

In the UWIFE survey region, there are thousands of small star forming regions
\citep{2002ARep...46..193A}.  
There are many objects related to the
outflow
phenomenon detected at various wavelength: 202 MHOs \citep[Molecular hydrogen
emission-line Objects:][]{2010AA...511A..24D}, 106 EGOs
\citep[Extended Green Objects in GLIMPSE
survey:][]{2008AJ....136.2391C}, 55 CO outflows
\citep{2004yCat..34260503W}, 239 6.7-Ghz methanol masers
\citep{2005AA...432..737P}, 57 Class I Methanol masers
\citep{2007ARep...51..519V}, 170 OH masers \citep{2004AA...423..209S}.
On the other hand, number of HH obejcts and alike are relatively small
in this region. Most of HH objects are concentrated in the region of
 100\arcdeg $<$ $l$ $<$ 220\arcdeg, and only several sources are reported in our survey region:
7 HH objects \citep[Herbig-Haro
Objects:][]{2000yCat.5104....0R} and 4 HBC sources \citep[Herbig-Bell
Catalog:][]{1988cels.book.....H}.
UWIFE-SYSOP (UWIFE-Searching for the Young Stellar Outflows Project)
is a project to find
IFOs associated with
YSOs. In this project, we will not only make the general catalog
of the IFOs but also study the statistics of outflows and their physical
quantities with the evolutionary stages of YSOs.

\subsection{Massive Star Forming Regions}
The formation of massive stars ($M\ga8\,M_{\sun}$) is still unclear in many aspects \citep{Zinnecker(2007)ARA&A_45_481}.
One question to be cleared is how massive stars obtain their mass.
It has been increasingly reported that, as in low-mass stars, the disk-mediated accretion process seems to be underway in forming massive stars \citep[e.g.][]{Beuther(2002)A&A_383_892,Wu(2004)A&A_426_503,SanJose-Garcia(2013)A&A_553_A125,Cooper(2013)MNRAS_430_1125}.
However, it is still uncertain if such a disk accretion works in the mass range of $M\ga25\,M_{\sun}$ \citep{Zinnecker(2007)ARA&A_45_481}.
One way to grasp the accretion process is tracing the outflow features, because the outflow process is closely related with the disk accretion process (cf.~section \ref{sec:discussion-yso}).
In the following sections, we describe how UWIFE will contribute to the elucidation of massive star formation through the jet and outflow phenomena.

\subsubsection{Outflow Features in Infrared Dark Clouds}

Infrared dark clouds (IRDCs), 
seen silhouette against the bright Galactic background in mid-infrared (MIR), 
are cold ($< 25$ K) and very dense ($n_{\rm H_2} > 10^5 {\rm cm}^{-3}$) 
interstellar clouds with high column densities
($\sim 10^{23} - 10^{25} {\rm cm}^{-2}$; \citet[][and references
therein]{2006ApJ...639..227S}) 
so that IRDCs are believed to be a probable sites 
where massive stars are forming.
Due to large extinction, 
IRDCs have usually been studied at longer wavelengths
(e.g., far-infrared or millimeter) to investigate prestellar or  
starless cores in IRDCs and their filamentary structures
\citep[e.g.,][]{2006ApJ...641..389R, 2010ApJ...719L.185J, 2012MNRAS.422.1071W}.
Although most IRDCs or their cores ($>50 \%$) do not show 
a signature of on-going star formation in MIR, 
there are some cores with bright MIR stellar sources
indicating star formation activity \citep{2009ApJS..181..360C}. 
These star-forming cores are, however, still deeply embedded in a cloud 
with large extinction, and few studies are made toward IRDCs for searching for
outflow features in NIR. 
Therefore, it is worthwhile to examine the data from 
an unbiased \feii\ survey data 
to detect outflows in IRDCs, particularly in rather evolved IRDCs which show 
active star formation for understanding the process of high-mass star formation 
in an early evolutionary phase.

\begin{figure*}
\epsscale{0.9}
\plottwo{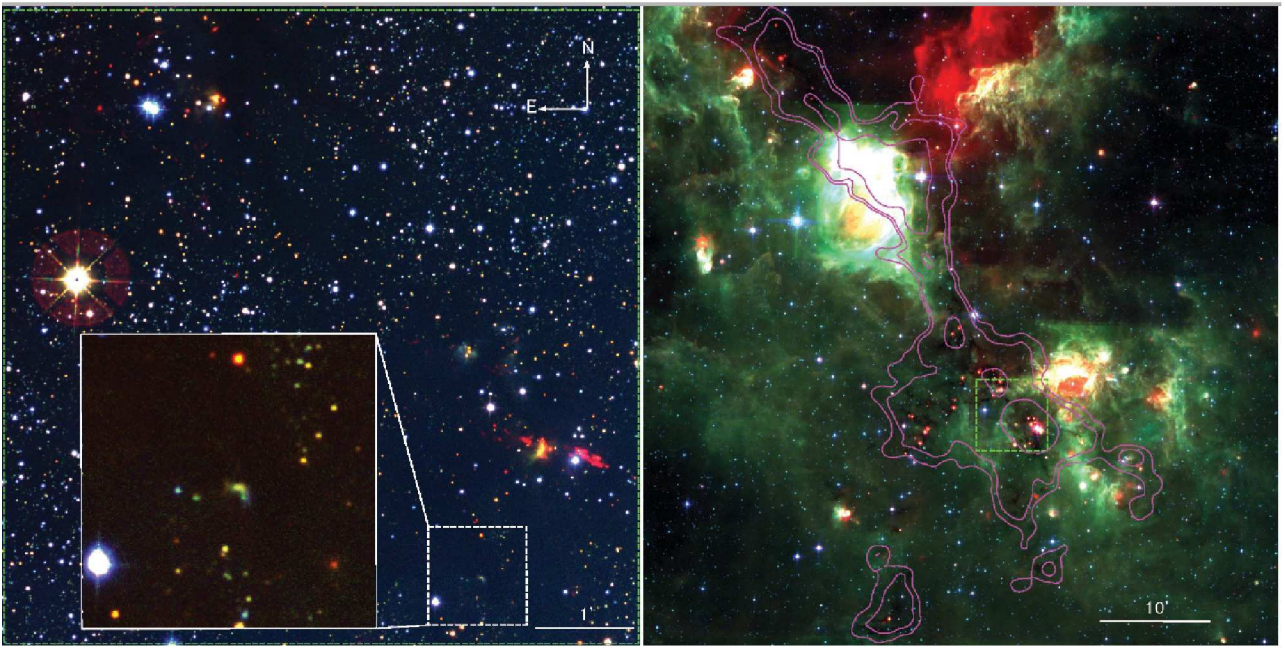}
\caption{(Left) Three-color composite image 
of the center of the IRDC G53.2 
made by UKIRT/WFCAM \Htwo(R)+[Fe II](G)+J(B) from UWISH2+UWIFE+GPS.
The presented region in the whole IRDC is marked by green-dashed box
in the right figure. The inset presents [Fe II] emission identified 
in this area. (Right) Three-color composite image of the whole IRDC 
G53.2 made by Spitzer IRAC/MIPS
24\,\um(R)+8.0\,\um(G)+5.8\,\um(B).
GRS $^{13}$CO $J=$1--0 contours are overlaid in magenta. 
The outermost contour presents the boundary of the IRDC.
\label{fig:irdc}}
\end{figure*}

One such cloud is the IRDC located at $(l, b) \sim (53.2,\, 0.0)$ which
shows a number of bright MIR stellar sources, likely YSOs, 
along its long, filamentary structure; here, we refer to it as ``IRDC G53.2.''.
% Kim et al. (in prep.) found that 
% the IRDC G53.2, extending $\sim 30$ pc, 
% is very well coincident with a CO cloud 
% at $v \sim 23.5$ \kms (or at $d \sim 1.7$ kpc) 
% in the Galactic Ring Survey (GRS) data \citep{2006ApJS..163..145J} 
% as presented by contours in Figure~\ref{fig:irdc}. 
The UWISH2 survey data reveals ubiquitous outflow features
around YSO candidates in the IRDC G53.2 (Figure~\ref{fig:irdc}, left).
As noted in other recent studies \citep[e.g.,][]{2009A&A...496..153D}, %(e.g. Davis et al. 2009),
the \Htwo\ outflows are particularly identified around earlier classes of YSOs 
(e.g., Class I and Flat Class) which contribute $\sim 50\%$ of 
the YSO candidates in the IRDC G53.2 
based on the analysis of {\it Spitzer} data (Kim et al. in prep.).
In the UWIFE image of the IRDC G53.2,
only a few \feii\ emissions have been detected so far.
The inset in Figure~\ref{fig:irdc} (left) shows \feii\ emission
identified around a YSO candidate in the IRDC.
A point-like knot shows \Htwo\ emission as well but with a little 
shifted peak between the \feii\ and \Htwo\ emissions.
A preliminary analysis suggests that 
the number of identified \feii\ emissions is much smaller 
compared to that of \Htwo\ emissions in the same area.

% As mentioned above, 
% few observations of \feii\ outflows/jets have been
% conducted for massive star forming regions such as IRDCs.
% The UWIFE survey, together with UWISH2, will help us to study 
% the outflow/disk systems around high-mass protostars 
% in an early evolutionary phase
% and investigate the currently proposed theories on 
% the massive star formation.
Investigation 
of the characteristics of 
\feii\ emissions associated with YSOs in IRDCs 
in the entire UWIFE survey area, where $\sim$ 300 IRDCs 
which have GRS counterparts \citep{2006ApJ...653.1325S} are included,
will provide much insight on the nature of jets and outflows of YSOs.
In particular, comparative studies
between the IRDCs and the low-mass star forming regions will be of
importance.
We note that there already are on-going or planned UWISH2 projects 
to search for outflows/jets toward massive star forming regions 
or filamentary clouds in \Htwo\ \citep{2011MNRAS.413..480F,
2012MNRAS.421.3257I,2012ApJS..200....2L}.
The UWIFE survey, therefore, will be a good 
complimentary data to expand our understanding to the physical 
structure of the outflows of high-mass protostars.

\subsubsection{Outflow Features around Ultracompact \ion{H}{2} regions}
\label{sec:discussion-uchii}
An ultracompact \ion{H}{2} region (UCHII) indicates a very compact \ion{H}{2} region, whose typical size, density, and emission measures are $\la10^{17}$ cm, $\ga10^4$ cm$^{-3}$, and $10^7$ pc cm$^{-6}$, respectively \citep{Churchwell(2002)ARA&A_40_27}.  
It is thought to be a ``late'' stage of massive young stellar objects (MYSOs), no longer accreting significant mass \citep{Churchwell(2002)ARA&A_40_27,Zinnecker(2007)ARA&A_45_481}.
More specifically, it is the period between the rapid accretion phase when the central object is being formed and the ionized phase when the larger, more diffuse, less obscured \ion{H}{2} region is being produced.
UCHII's unique position in the MYSO evolutionary track enables us to study the history of accretion processes as follows. 

Since UCHII's natal clumps are not completely destroyed yet, one may expect that the materials ejected during the prior, active accretion phase produce shocked outflow features around UCHIIs.
By tracing these ``footprint'' outflow features around UCHIIs, we can study the history of accretion process, because the outflow process is closely related with the accretion process (cf.~section \ref{sec:discussion-yso}).
These features can be observed through radiative cooling lines, such as \feii\ 1.64 \um, H$_2$ 2.12 \um, and CO radio lines \citep{Hollenbach(1989)ApJ_342_306,Neufeld(1989)ApJ_344_251,Kaufman(1996)ApJ_456_611,Wilgenbus(2000)A&A_356_1010,Flower(2010)MNRAS_406_1745}. 
Indeed, CO outflow features have been observed around UCHII regions, such as G5.89-0.39 \citep{Watson(2007)ApJ_657_318,Wood(1989)ApJS_69_831} and G18.67+0.03 \citep{Cyganowski(2012)ApJ_760_L20}.
The dynamical timescale of these CO outflow features is $>10^3$ yr, which is comparable or greater than the typical lifetime of UCHIIs \citep[$\ga5\times10^4$ yr,][]{Wood(1989)ApJS_69_831,Gonzalez-Aviles(2005)ApJ_621_359} and MYSO jet-phase \citep[$\sim4\times10^4$ yr,][]{Guzman(2012)ApJ_753_51}.

% However, the CO outflow observations were performed with low spatial
% resolutions greater than several arc seconds, limiting detailed
% morphological study of the outflow features.
% \feii\ emission features, thanks to its high spatial
% resolution, help us also to understand the multiplicity 
% of star formation better when compared to CO surveys at 
% low spatial resolution.  Aligned \feii\ features enable 
% us to identify the location of the YSO more precisely.  
% This is important in massive star formation, where it 
% always happens in clusters.  Especially in the case of 
% UCHIIs where it is debated if they still accumulate 
% mass by accretion, identification of aligned \feii\
% emission features will enable us to understand if the 
% \feii\ jet is from the UCHII or some other younger 
% source near it.  Even if we detect CO outflows ``from'' a 
% UCHII, unless it is from an interferometric observation
% with high angular resolution, we cannot say for sure if 
% the UCHII is responsible for it.  However, from radio 
% continuum studies using interferometers, we know the 
% location of the UCHII source with far higher accuracy. 
% Therefore, a high angular resolution jet/outflow tracer 
% will enable us to say with more confidence if the UCHII 
% is still accreting, or the outflow is from a younger 
% sources in the field of view.

\begin{figure}
\epsscale{1.}
\plotone{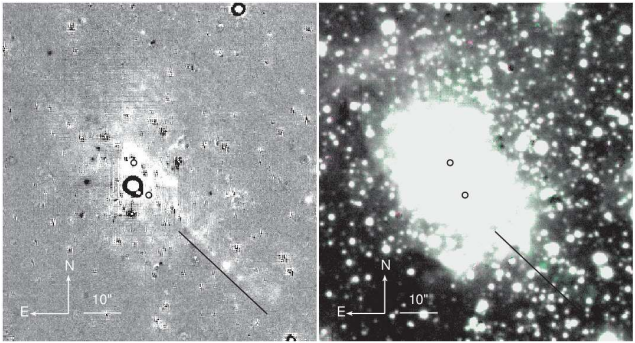}
\caption{\feii\ outflow feature around a UCHII region
  G025.3824$-$00.1812 (\emph{upper-circle}) and G025.3809$-$00.1815 (\emph{lower-circle}) of \citet{Purcell(2013)ApJS_205_1}: (\emph{left})
  Continuum-subtracted \feii\ image and (\emph{right}) RGB color-composite image (R=H, G=\feii, B=H). The black line indicates the outflow features.
\label{fig:uchii}}
\end{figure}

A search for \feii\ features around UCHIIs are under way using a UCHII catalog.
The catalog is from the CORNISH survey \citep{Hoare(2012)PASP_124_939,Purcell(2013)ApJS_205_1}, which covers the same part of the Galactic plane as UWIFE.
Shinn et al. (in prep.) have found \feii\ outflow features around several UCHIIs. 
One example, G025.3824$-$00.1812 and G025.3809$-$00.1815, is shown in Fig.~\ref{fig:uchii}, where the southwestward \feii\ outflow/jet feature is evident.
The outflow feature seems to emerge from the direction towards the two UCHIIs and has a well-collimated appearance.
The \feii\ features around UCHIIs detected from the unbiased UWIFE survey enable us to extract some general properties of the MYSO accretion process.
The outflow morphology, the outflow mass loss rate, and their relation with the MYSO physical parameters would be the initial outcomes
\citep[cf.][]{Shinn(2013)ApJ_777_45}.  
Based on those results, we may investigate if the outflow features support the disk accretion process, and may infer the final stellar mass where the disk accretion seems to work.

\subsection{Nebulae around Evolved Stars}
\label{sec:discussion-evolved}

% The mass of a star is the key parameter that governs almost every
% aspect of the stellar evolution. Stellar masses vary throughout the life
% of stars and mass--loss is the dominating factor in their late stage
% of evolution. Unfortunately, the mass loss rate from stars still 
% remains to be a highly uncertain parameter. 
% For low- and intermediate mass stars (initial
% mass $\lesssim$ 8 \msun), 
% mass-loss is expected to occur primarily via a 
% slow wind with a high mass-loss rate during their
% asymptotic giant branch (AGB). More massive stars go through various
% evolutionary stages (red super giant (RSG), luminous blue variable (LBV)
% and Wolf-Rayet (WR) stars).
% The mass-loss mechanism varies for each evolutionary stage, 
% and episodic and eruptive mass loss may play a critical role 
% for stars close to the Eddington-limit like LBV stars.

Circumetellar nebulae around evolved stars are fossil record of their
mass-loss history. They often represent dense structure resulting
from sudden changes in wind characteristics.  These circumstellar
nebulae are often known from their optical emission lines, but
infrared emission from cold and dense medium is becoming more
important. In principle, emission in different regimes provide us 
with a more
complete view of the progenitor star system and its evolution.

\subsubsection{Planetary Nebulae}
\label{sec:discussion-pne}

Planetary nebulae (PNe) are the most frequent type of nebula around evolved
stars.  
The nebula consists of material ejected during the AGB
phase and ionized by the UV radiation from the newly-formed hot central star
\citep{2010PASA...27..129F}.  
% The wind during the AGB phase is
% quasi-isotropic, but the wind during the PN phase is
% nonspherical, often highly collimated, which gives a stunning variety of
% PN shapes. The physical mechanisms responsible for this have been the
% subject of debate \citep[e.g.,][]{2011apn5.confE....Z}.
%
More than a thousand PNe are currently known in our Galaxy
\citep[e.g.,][]{1992secg.book.....A}. However, most of the known ones are 
high above the Galactic plane and the number of PNe along the Galactic
plane is relatively small.  
With the advent of sensitive \Halpha\ survey of the
Galactic plane such as IPHAS \citep{2005MNRAS.362..753D} and VPHAS+, a
large number of PNe in the plane is being revealed
\citep{2009A&A...502..113V,2009A&A...504..291V,2010PASA...27..166S}.
Furthermore, the NIR line surveys like UWISH2 and UWIFE have the
potential to uncover a significant population PNe along the Galactic
plane. These surveys will help us to constrain their space density and
any variation along the plane which is related to various routes 
of PN formation. Also the findings of numerous new PNe
will enable more rigorous statistical
analysis.

PNe are often bright in the \Htwo\ 2.122 \um\ line.
\citet{1996ApJ...462..777K}
detected \Htwo\ emission from
23 PNe
out of 60 PNe in their narrow-band imaging survey of 2.122 \um\ emission.
The 2.122 \um\ \Htwo\ emission is most prominent in PNe of bipolar
morphology and they suggested a physical connection.  On the other hand,
surveys of PNe in \feii\ 1.644 \um\ have been sparse.
\citet{1999ApJS..124..195H} observed a sample of PNe with
medium-resolution (R$\sim$700) near-infrared ($\lambda = 1 - 2.5 \um$)
spectra. Among 41 they observed, 16 showed \Htwo\ 2.122 \um\ emission but
only 3 showed \feii\ 1.644 \um\ emission. While the fraction of PNe with
detected \feii\ 1.644 \um\ emission was small, the emission has been 
useful in studying the nature
of mass loss, as demonstrated in the case of M2-9 \citep{2005AJ....130..853S}.

\begin{figure}
\epsscale{1.}
\plotone{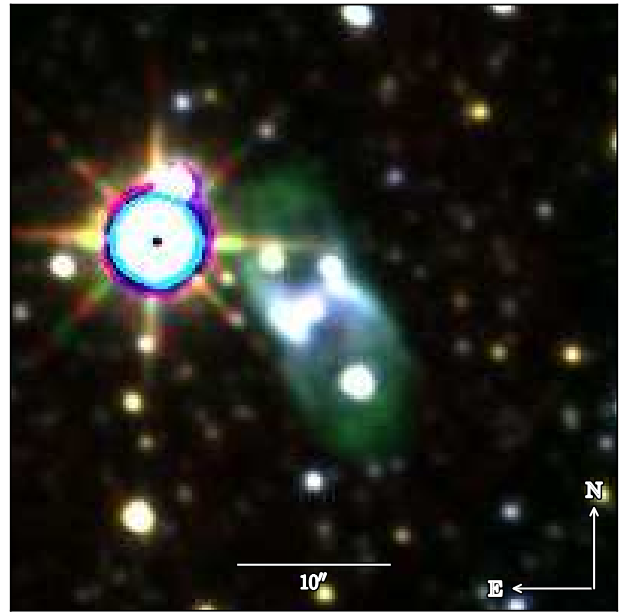}
\caption{RGB composite image of PN M 1-51.
The color scheme is same as in Figure~\ref{fig:gthirtyfive}.
This is one of the PN showing \feii\ emission without \Htwo.
\label{fig:pn}}
\end{figure}

The initial search of \feii\ emission from the known PNe in the UWIFE survey
resulted in a half dozen of detection from around 29 known PNe (25
from \citet{1992secg.book.....A} and 4 from \uwishtwopaper).  
Most of these PNe with \feii\ emission have accompanying \Htwo\ emission.
Nonetheless, 3 of them are not detected in \Htwo\ (e.g., PN M 1-51 as in
Figure~\ref{fig:pn}). For comparison, 
\citet{1996ApJ...462..777K} found all the 3 PNe detected in \feii\ 
show accompanying \Htwo\ emission.  
Study of PNe connecting their emission characteristics to their
morphology will be important as it may reveal how the mass--loss history of
progenitors affects the shaping of PNe.

\subsubsection{Nebulae around Evolved Massive Stars}
\label{sec:discussion-lbv}

Mass loss in massive stars plays a more critical role than in less
massive ones.  
% An oversimplified picture of the evolution of the
% massive stars (massive enough to undergo core-collapse eventually) is
% that stars of 8 -- 25 \msun\ explode as red or blue supergiant
% (RSG/BSG) stars with thick hydrogen envelopes while more massive stars
% (M $\gtrsim$ 25 \msun) end their lives as Wolf-Rayet (WR) stars (i.e.,
% massive stars which have lost most or all of 
% their hydrogen envelope through stellar winds).  For stars of M
% $\lesssim$ 25 \msun, dense, slow and steady RSG winds seem to dominate
% the overall mass loss, while there is evidence for the episodes of
% strongly enhanced mass loss for extreme red supergiants like VY CMa
% \citep[e.g.,][]{2009AJ....137.3558S}.  For more massive stars, 
% both line-driven winds and eruptive mass loss may  play an important 
% role for the formation of WR stars. 
%
An important class of evolved massive stars are luminous blue variables
(LBVs). LBVs 
belong to the most luminous stars in our Galaxy 
(thus with very large initial masses)
typified by their irregular variabilities, which are sometimes
associated with eruptive mass loss. The best example of LBV star is
$\eta$ Car which is surrounded by an extremely massive circumstellar shell
($M\sim15⁢\msun$), believed to be a product of its 1840's great
ejection. 
The line emission from circumstellar shells around some LBVs
has been a valuable tool to 
understand not only
the physical process responsible for LBV mass loss
but also the nature of
the central star.
NIR lines, such as \feii, are well suited to study the shells around 
evolved massive stars as they are less affected by interstellar extinction.
\feii\ emission is so far found in a half dozen of LBVs
\citep[][]{2002MNRAS.336L..22S}, 
and detailed studies of $\eta$~Car have demonstrated the usefulness
of the \feii\ emission \citep{2006ApJ...644.1151S}.

\begin{figure}
\epsscale{1.}
\plotone{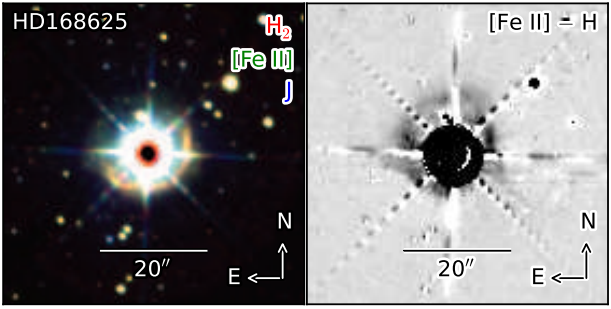}
\caption{(\emph{left}) RGB colour composite image of LBV HD168625.
Red, green and blue each represents \Htwo\ from the UWISH2, 
\feii\ from the UWIFE and broad--band J from the GPS survey.
(\emph{right}) Continuum--subtracted \feii\ image of HD168625. 
\label{fig:lbv}}
\end{figure}

In our preliminary study, we searched for \feii\ emission
associated with known LBVs and candidate LBVs from \citet{2005A&A...435..239C}. 
Among several sources covered by the UWIFE survey, we detected \feii\ emission
from a few of them. 
An example is HD~168625 as shown in Fig.~\ref{fig:lbv}.
HD~168625 is an intriguing system showing a bipolar nebula several times 
larger than its equatorial dust torus \citep{2007AJ....133.1034S}, 
quite similar to what is found around 
SN~1987A. We found both \Htwo\ and \feii\ emission around the
equatorial torus, but with different morphology. The morphology 
of \Htwo\ emission is similar to \halpha\ and the thermal emission 
from the dust \citep{2002AJ....124.1625P} and is sharply
peaked along the torus in the southwestern side. On the other hand, 
the \feii\ emission is more diffuse and more extended toward 
the northeastern side. 
The only other LBV that shows both \Htwo\ and \feii\ emission 
is $\eta$~Car. Follow--up spectroscopic observations of
HD~168625 will reveal detailed mass--loss history of the central star, 
as in the case of $\eta$~Car \citep{2006ApJ...644.1151S}.

\subsection{Supernova remnants}
\label{sec:discussion-snrs}

%Supernova remnants (SNRs), which are expanding gaseous remnants
%of supernova (SN) explosions, are by nature strong \feii--line emitters.
When a SN explodes, a strong shock with a typical
expansion velocity of $\sim$ 10,000 \kms\ is produced,
which later decelerates
and becomes radiative at the speed of a few hundreds \kms.
Along with this front shock expanding into the surrounding medium,
there exists a reverse shock which
propagates back into SN material.
Strong NIR \feii\ lines are emitted when these shocks interact
with dense material, which could be either the SN or
the circumstellar material for young SNRs, and usually interstellar material
for middle-aged SNRs 
\citep[e.g.,][see also Koo~2013 and references therein]{1989A&A...214..307O,
1990A&A...240..453O,
2007ApJ...657..308K,
2009ApJ...691.1042L,
2009ApJ...703L..81M,
2013ApJ...770..143L}.
%The relatively low excitation energies ($\le 1.3\times 10^4$~K) 
%of the ground terms of the Fe$^+$ ion make the near-IR \feii\ lines
%an excellent tracer of the cooling gas in the
%downstream where Fe is mostly in Fe$^+$ while hydrogen is only partially
%ionized 
%\citep{1989A&A...214..307O,
%1990A&A...240..453O,
%2013arXiv1304.3882K}.
Moreover, Fe, being the end product of the stellar 
nucleosynthesis process, forms a main component of 
the supernova ejecta; the materials ejected from the 
deep layers of the progenitor are often enriched in Fe.
The Fe abundance can also be significantly
enhanced in a shocked region if the interstellar dust is destroyed.

\begin{figure*}
\epsscale{1.}
\plottwo{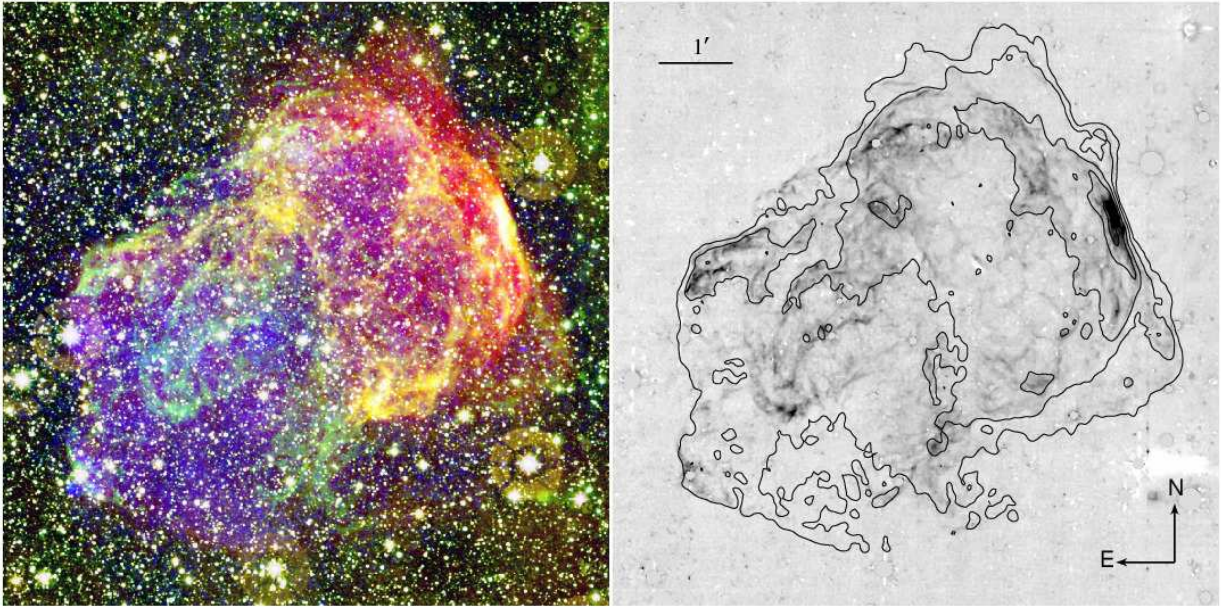}
\caption{
$Left$: Color composite figure of 3C~391,
composed of 1.4 GHz radio continuum (red), H$_2$ 2.122 $\mu$m (yellow), \feii\ 1.644 $\mu$m (green),
and Chandra X-ray (blue) images
\citep{2006AJ....131.2525H, 2011MNRAS.413..480F, 2004ApJ...616..885C}.
$Right$: Continuum-subtracted  \feii\ image.
Superimposed are VLA 1.4~GHz radio contours with
intensity levels of 2.0, 5.6, 17, 35, and 60 mJy beam$^{-1}$
and the beam size of $6 \farcs 2$ $\times$ $5 \farcs 4$.
\label{figsnr}}
\end{figure*}

In the survey area, there are over 70 known SNRs mostly identified by
radio and X-ray observations \citep{2009BASI...37...45G}.
We search for the \feii\ emission around the positions of the known SNRs.
Our preliminary results reveal that 
20--30 \% of known SNRs show \feii\ emission,
including previously observed \feii-emitting SNRs such as 
G11.2$-$0.3 \citep{2007ApJ...657..308K}, 
W~28 \citep{2005ApJ...618..297R}, 
G21.5$-$0.9 \citep{2012A&A...542A..12Z},
3C~391 \citep{2002ApJ...564..302R},
3C~396 \citep{2009ApJ...691.1042L},
W~44 \citep{2005ApJ...618..297R},
and
W~49B \citep{2007ApJ...654..938K}. 
In most cases, our \feii\ images provide high-quality, wide-area views
of detected SNRs, for which, sometimes, only parts of the area were
covered by the previous observations.  In addition, we find new
\feii-emitting SNRs.  Many of them show near-infrared H$_2$ emission
identified by UWISH2 survey and/or thermal X-ray
emission in the center, implying that they are located within dense
environments.  For example, Figure~\ref{figsnr} shows a color image of
3C~391, which is one of the bright \feii-emitting SNRs in the survey
area.

\subsection{Other Sources and Unbiased Search}
\label{sec:discussion-vriables}

\begin{figure*}
\epsscale{1.}
\plottwo{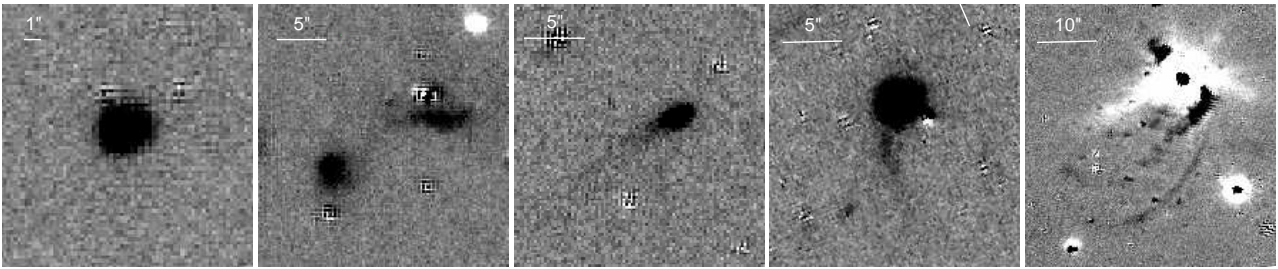}
\caption{Continuum--subtracted \feii\ images of IFOs of various 
morphologies in the
UWIFE survey area (from left to right): point-like, amorphous,
cometary, jet/ filamentary and shell-like. 
\label{fig:unidentified}}
\end{figure*}

The UWIFE survey is an {\em unbiased} survey, so that we have an
opportunity to detect ``every'' dense, shocked gas with bright \feii\ line
emission in the inner Galaxy, some of which would be serendipitous
discoveries. In order to identify IFOs in the UWIFE survey area,
we search for IFOs in the
continuum-subtracted \feii-line images visually using the naked
eye. At the time of writing, we have detected 111 IFOs in 10.5 deg$^2$
from l=53$\arcdeg$ to 60$\arcdeg$.  The detected sources include
YSO/YSO candidates, AGB/RCB (R Coronae Borealis stars) stars and their 
candidates, EGOs
(Extended Green Objects), and star forming regions.  But most ($\sim
85$\%) of them do not have counterparts in SIMBAD. The detected IFOs
are classified into point-like or diffuse based on their
morphology. The diffuse sources can be further divided into amorphous,
cometary, jet/filamentary, or shell--like. 
Figure~\ref{fig:unidentified} shows examples of each category.
To better systematize the unbiased search of IFOs, an automatic 
computer--aided method is being developed. We note, however, that
a significant number of point-like sources we detected above might be
variables instead of [Fe II]-line emitting sources.

\section{Concluding Remarks}
\label{sec:conclusion}

In this paper, we present an overview and initial outlook of 
the UKIRT Widefield Infrared Survey for \feii\ (UWIFE). The UWIFE
survey is an imaging survey of the first Galactic quadrant 
($7\arcdeg < l < 62\arcdeg$; $|b| \lesssim 1\fdg5$) using a narrow--band filter
centered on the \feii\ 1.644 \um\ emission line, which 
is a good tracer of dense shock-excited gas. The survey
started in the summer of 2012 and was completed as of 2013.
The resulting images have a median seeing of 0.8\arcsec\ with
a 5$\sigma$ limiting magnitude of $\sim$ 19$^{\mathrm{th}}$ mag and will provide
valuable resources to study the Galactic plane.

Our own Galaxy and its inner workings such as how new stars are born
and die are still mysterious. Unbiased imaging surveys of
the galactic plane in the infrared
broadband such as Glimpse and MIPSGAL have
greatly increased our understanding of the Galaxy.  A significant
advantage of our UWIFE survey is its complementarity with these
existing and/or upcoming surveys, in particular with the UWISH2 survey.  The
line emission such as the \feii\ 1.644 \um\ line 
and the molecular hydrogen 1-0 S(1) emission line at
2.122 \um\ will complement the broadband
surveys and probe dynamically active components of interstellar
medium. The high-spatial resolution narrow-band WFCAM
images from the UWIFE and UWISH2 surveys will pin--point regions of active
interaction in the complex environments. Furthermore, 
these imaging survey will provide wealth of potential targets for
follow--up spectroscopic observations.

\acknowledgements 

This work was supported by K-GMT Science Program 
funded through Korea GMT Project operated by Korea
Astronomy and Space Science Institute (KASI).  
H.-G.~Lee acknowledges support 
from a Grant-in-Aid for Japan Society of Promotion of Science (JSPS) fellows
(No. 23$\cdot$01322).
H.-J. Kim was supported
by NRF(National Research Foundation of Korea) Grant funded by the
Korean Government (NRF-2012-Fostering Core Leaders of the Future Basic
Science Program).
D. Moon was supported by Korean Federation of Science and Technology Societies (KOFST).
M.-G.~Lee, B.-C. Koo, H.-M.~Lee and M.~Ishiguro were
supported by the National Research Foundation of Korea
(NRF) grant funded by the Korea Government (MSIP) (No.
2012R1A4A1028713).

% \bibliography{ms,jhshinn2/references.bib}

\end{document}